\documentclass[aps,prd,amsfonts,nofootinbib,preprint]{revtex4}

\usepackage{graphics,epsfig}

\begin{document}
\preprint{WM-03-108}
\title{\vspace*{0.5in} Positive Parity Pentaquarks Pragmatically Predicted
\vskip 0.1in}
\author{Carl E. Carlson} \email[]{carlson@physics.wm.edu}
\author{Christopher D. Carone}\email[]{carone@physics.wm.edu}
\author{Herry J. Kwee}\email[]{herry@camelot.physics.wm.edu}
\author{Vahagn Nazaryan}\email[]{vrnaza@wm.edu} 
\affiliation{Nuclear and Particle Theory Group, Department of Physics,
College of William and Mary, Williamsburg, VA 23187-8795}
\date{October 2003}
\begin{abstract}
We consider the possibility that the lightest pentaquark is a parity
even state, with one unit of orbital angular momentum. Working within
the framework of a constituent quark model, we show that dominant
spin-flavor interactions render certain parity-even states lighter than
any pentaquark with all quarks in the spatial ground state. For such
states, we focus on predicting the mass and decays of other members of
the same SU(3) flavor multiplet. Specifically, we consider the
strangeness $-2$ cascade pentaquarks, which are relatively immune to
mixing. We take into account flavor SU(3) breaking effects originating
from the strange quark mass as well as from the structure of the
spin-flavor exchange interactions themselves.  We predict the lightest
cascade pentaquarks at approximately $1906$~MeV, with a full width
$\sim 3$ times larger than that of the $\Theta^+$.
\end{abstract}
\pacs{}
\maketitle

\section{Introduction}\label{sec:intro}

The existence of an exotic baryon state containing an antiquark in its lowest Fock component has been verified by the observations at a number of laboratories of a strangeness $+1$ baryon at $1540$~MeV with a narrow width~\cite{nakano,barmin,stepanyan,barth,kubarovsky,asratyan}.  In distinction to all previously discovered baryons, such a state must have four quarks and an antiquark in its minimal Fock component.  The present  example, which has quark content $udud \bar s$, was known as $Z^+$ during its advent, and now seems generally called $\Theta^+$  (e.g., ~\cite{stepanyan,barth,kubarovsky,asratyan}).

There are a number of pre-discovery theoretical studies of pentaquarks~\cite{hogaasen,strottman,roiesnel,oh,diakonov}, some including heavy quarks in the pentaquark state~\cite{heavyexamples,leandri}.  Of particular note is~\cite{diakonov}, which, though it has been criticized~\cite{weigel,coheneinz}, advanced the field by predicting in the context of a chiral soliton model a narrow pentaquark only 10 MeV away from the discovery mass.  Since the $\Theta^+$ discovery, there has been a flurry of papers studying pentaquark properties in constituent quark models~\cite{cqm,stancu,jaffew,us,jennings}, other aspects of pentaquarks in soliton models~\cite{coheneinz,csm}, production of pentaquarks, including in heavy ion collisions~\cite{hi}, non-observance of pentaquarks in earlier hadronic experiments~\cite{hr}, pentaquarks in the large $N_c$ limit~\cite{cohenzwei}, and other pentaquark topics~\cite{leftovers,sugiyama}.

At present, the spin and parity of the $\Theta^+$ are experimentally unknown.  A majority of the theoretical papers, including all the chiral soliton papers, treat the state as positive parity.  A minority, including an earlier work by the present authors~\cite{us}, have considered the possibility of negative parity~\cite{sugiyama}.  All theory papers, to our knowledge, consider the $\Theta^+$ to be spin-1/2.  Regarding the isospin, a $\Theta^{++}$ signal has been sought and not found~\cite{barth}, so that the $\Theta^+$ appears to be isoscalar and hence a member of a pentaquark flavor antidecuplet.

In the present work, we focus on understanding how a positive parity
state could emerge as the lightest pentaquark, in the context of a
constituent quark model~\cite{stancu,glozman,jennings}.  We explore the consequences of the ensuing
picture for other states in the pentaquark antidecuplet. Positive
parity pentaquarks in a constituent quark model require a
negative-parity spatial wave function, obtained by putting one quark in
the lowest P-state of a suitable collective potential. One could
entertain more complicated excited state scenarios also
(e.g.,~\cite{jaffew}).  Here we discuss a plausible mechanism that
changes the level ordering so that a state with an excited wave
function becomes the lightest one.  In this approach, the positive
parity of the state is a consequence of the quark-quark pairwise
potential and the chosen symmetry structure of the flavor-spin wave
function.

Insight comes from studies of three-quark baryons~\cite{gr}, where the
level ordering of the first excited positive and negative parity states
is reproduced correctly in an effective theory where the dominant
pairwise interaction is flavor-spin dependent. One-gluon exchange gives
only a color-spin dependent force. Flavor-spin dependent interactions
can be pictured as arising from the interchange of quark-antiquark
pairs with the quantum numbers of pseudoscalar mesons.  However, the
effective theory viewpoint does not require that one commit to a
specific model for the underlying physics.  Skyrmion or instanton induced interactions could be described equally well by the effective field theory introduced below.

In the next section, we demonstrate how effective flavor-spin interactions lead to the correct $q^3$ mass spectrum, and in particular rectify the level order of the Roper and negative parity resonances.  We also discuss semiquantitatively the consequences of the flavor-spin interaction for the pentaquark system.   Section~\ref{sec:fp} includes a more detailed numerical analysis, taking into account the breaking of SU$(3)_F$ symmetry.  We give predictions which are new in the effective theory context for the mass and decays widths of other members of the pentaquark antidecuplet, particularly the exotic cascade states $\Xi_5$. 
In a constituent quark model with flavor independent spin-splittings, the difference between the $\Xi_5$ and $\Theta^+$ masses is just that obtaining from an additional strange quark, about 150 MeV~\cite{us,jaffew}.  We find that the flavor symmetry breaking stretches out this mass gap considerably, pushing the $\Xi_5$ mass to about 1900 MeV.  This is nonetheless much smaller than the mass gap predicted in the chiral soliton model in~\cite{diakonov}.  The predicted width of a 1900 MeV $\Xi_5$ is still narrow, which suggests that the $\Xi_5$ should be distinguishable from background.

\section{Framework}\label{sec:fw}

A key feature of the flavor-spin interaction is that it is most
attractive for states that have the most symmetric flavor-spin wave
functions.  If the interaction has
exact SU$(3)_F$ flavor symmetry (which may not be the case and which we do not
assume later), then the mass shift is given by
\begin{equation}					\label{eq:one}
\Delta M_\chi = -C_\chi \sum_{\alpha<\beta} 
	\left( \lambda_F \sigma \right)_\alpha  \cdot
	\left( \lambda_F \sigma \right)_\beta  \ ,
\end{equation}

\noindent where the sum is over all $qq$ and $q \bar q$ pairs
$(\alpha,\beta)$, the $\vec\sigma_\alpha$ are Pauli spin matrices for
quark or antiquark $\alpha$, and $\vec \lambda_{F\alpha}$ are flavor
Gell-Mann matrices. Coefficient $C_\chi$ is a positive number.  Let us
focus on states or components of states that contain quarks only.  If
the flavor-spin state is symmetric overall, then one may write the wave
function as a sum of terms in which a given pair of quarks is singled
out and in which the individual spin and flavor wave functions of the
given pair are either both symmetric or both antisymmetric. In either
case, the expectation values of $\vec
\sigma_\alpha\cdot\vec\sigma_\beta$ and
$\vec\lambda_{F\alpha}\cdot\vec\lambda_{F\beta}$ for that pair have the same sign and yield maximal attraction.

The most significant contribution to Eq.~(\ref{eq:one}) in a pentaquark
state comes from the sum over the $q^4$ component. Let us compare the
situation of four quarks in S-states [$S^4$] to one where one quark is
in a P-state and three are in S-states [$S^3 P$].  The color state of the $q^4$ must be a $\mathbf 3$, which for four quarks is a mixed symmetry state.  If all quarks are in the same spatial state, then of necessity the flavor-spin state must also be of mixed symmetry.  However, for the $S^3P$ combination, one can have a mixed-symmetry spatial state and a color-orbital state that is totally antisymmetric.  The flavor-spin wave function is then totally symmetric, and leads to the most attractive possible flavor-spin interaction.
We will compute below the numerical lowering of the $S^3P$ binding energy relative to the $S^4$, and show that it is dramatically large, more than enough to balance the extra energy associated with the orbital excitation.  This gives a semiquantitative understanding of the numerical results that we present in section~\ref{sec:fp}.

It is useful to recall how flavor-spin interactions work in the ordinary $q^3$ baryon systems, both to motivate our framework and to estimate numerical values for the parameters involved.  The dramatic problem that is solved is the level ordering of the $N^*(1440)$, the positive parity S-state excitation of the nucleon also known as the Roper resonance, and the $N^*(1535)$, the lightest spin-1/2 negative parity resonance, which we refer to as the $S_{11}$.

In the Bag model and in linear or harmonic oscillator confining potentials,  the first excited S-state lies above the lowest P-state, making the predicted Roper mass heavier than the lightest negative parity baryon mass.  Pairwise spin-dependent interactions must reverse the level ordering.  As mentioned earlier, color-spin interactions fail in this regard~\cite{friends},  while flavor-spin interactions produce the desired effect.  Since the $q^3$ color wave function is antisymmetric, the flavor-spin-orbital wave function is totally symmetric.  For all quarks in an S-state, the flavor-spin wave function is totally symmetric all by itself and leads to the most attractive flavor-spin interaction.  If one quark is in a P-state, the orbital wave function is mixed symmetry and so is the flavor-spin wave function, and the flavor-spin interaction is a less attractive .  In the SU$(3)_F$ symmetric case, Eq.~(\ref{eq:one}), one obtains mass splittings
\begin{equation}
\Delta M_\chi = \left\{ \begin{array}{cl}
	-14 C_\chi & \quad N(939), N^*(1440) \\
	 -4 C_\chi & \quad \Delta(1232) \\
	 -2 C_\chi & \quad N^*(1535)
\end{array} \right.   \ .
\end{equation}

\noindent Here we have approximated the $N^*(1535)$ as a state with total quark spin-1/2.

\begin{figure}

\centerline{ \epsfysize 3in \epsfbox{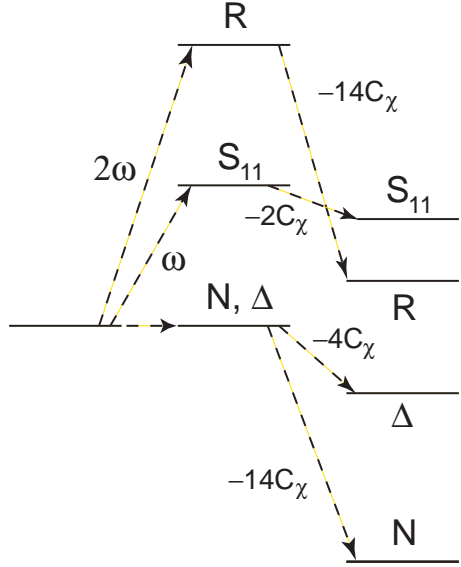}  }

\caption{Schematic view of the level reversal of the P-state and excited S-state for 3-quark baryons.}

\label{fig:spectrum}

\end{figure}

The scenario is shown in Fig.~\ref{fig:spectrum}.  Relative to some base mass, one first has the 2S--1S and 1P--1S splittings for the Roper and the $S_{11}$.  Then the flavor-spin pairwise interactions further split the spectrum into its final form, placing the Roper below the mass of the negative parity baryon.  We have worked with a small number of states to illustrate clearly how the mechanism works.  More extensive evidence that flavor-spin splitting is significant in the baryon spectrum is found in~\cite{gr,cclg,hael,carlos,myhrer}.


Returning to pentaquarks, the presence of a P-state now allows for a more rather than a less symmetric $q^4$ flavor-spin wave function.  The net result is that pentaquarks with $S^3 P$ four-quark components are lighter than the corresponding states with all quarks in the ground state.  One can estimate the advantage of this configuration as follows.  For the $q^4$ part of the state, the mass splitting of Eq.~(\ref{eq:one}) evaluates to,
\begin{equation}
\Delta M_\chi = - C_\chi \left\{ 4 C_6(R) - 8 N 
	- {4\over 3} S^2 - 2 F^2 \right\}  \ ,
\end{equation}

\noindent where $C_6(R)$ is the quadratic Casimir of the $SU(6)$ flavor-spin representation $R$, $N$ is the number of quarks, and $S^2$ and $F^2$ are the spin and flavor quadratic Casimirs of the state. (We normalize generators $\Lambda_A$ so that Tr $\Lambda_A\Lambda_B = (1/2) \delta_{AB}$.  A representation $R$ can be specified by its Young diagram, and a useful expression for the quadratic Casimir of representations of $SU(Q)$ is found in~\cite{djm95},
\begin{equation}
C_Q(R) = {1\over 2} \left( NQ - {N^2\over Q} + \sum r_i^2 - \sum c_i^2
	\right)
\end{equation}

\noindent where $r_i$ is the number of boxes in the $i^{th}$ row of the Young diagram, $c_i$ is the number of boxes in the $i^{th}$ column, and $N$ is the total number of boxes.)  For the present situation,
\begin{equation}
\Delta M_\chi = \left\{ \begin{array}{cl}
	-{28\over 3} C_\chi & \quad S^4 \\
	-28 C_\chi          & \quad S^3 P
	\end{array} \right.  \ .
\end{equation}

\noindent To make a $\Theta^+$, all four quarks are non-strange and the state is isospin-0.  Fermi symmetry requires the $S^4$ state to be spin-1.  The $S^3 P$ state can be spin-0, and we take it so.
Thus
\begin{equation}						\label{eq:six}
M(S^3 P) - M(S^4) = \hbar\omega - {56\over 3} C_\chi 
	\approx -310 {\rm\ MeV}  \ .
\end{equation}

\noindent For the numerical evaluation of Eq.~(\ref{eq:six}), we have assumed the $1P$--$1S$ level splitting of a harmonic oscillator potential, with $2\hbar\omega$ estimated from the nucleon-Roper mass difference; the coefficient $C_\chi$ is fixed by the nucleon-$\Delta(1232)$ mass splitting.  Adding the strange antiquark to the spin-0 $S^3 P$ state gives no further spin-dependent mass shift.  Adding the $\bar s$ to the spin-1 $S^4$ state does give a spin-dependent splitting can lower the mass, but not decisively.   Thus, the pentaquark state with an $S^3 P$ four-quark state is the lightest by a wide margin.

A key concern is the location of the other pentaquark states.  Particularly interesting are the other exotic members of the pentaquark antidecuplet, namely the isospin-3/2 pentaquark $\Xi_5$, or cascade, states.  To more accurately predict the masses and widths of these strangeness $-2$ states, or of other states of varying flavor, we should consider the effects of flavor symmetry breaking in the flavor-spin interaction.  Certainly one knows that isolated quark-antiquark pairs bind into states with flavor-dependent masses.  With flavor symmetry breaking we write the isospin-conserving, spin-dependent interaction as 
\begin{equation}							\label{eq:seven}
\Delta M = -C_{SI} \sum_{\alpha < \beta} 
		(\tau\sigma)_\alpha \cdot (\tau\sigma)_\beta
	-C_{47} \sum_{ \alpha < \beta, i = 4}^7 
		(\lambda^i\sigma)_\alpha \cdot (\lambda^i\sigma)_\beta
	-C_8 \sum_{ \alpha < \beta} 
		(\lambda^8\sigma)_\alpha \cdot (\lambda^8\sigma)_\beta  \ .
\end{equation}

\noindent The $\tau^i_{\alpha}$ are the isospin matrices for quark $\alpha$, the same as $\lambda^i_{\alpha}$ for $i=1,2,3$.  We find the coefficients by studying the mass splitting in the three-quark sector, as is reported in the next section.  Matrix elements of Eq.~(\ref{eq:seven}) in the pentaquark states (summing over all 5 constituents) are also presented, so that the splittings within the pentaquark antidecuplet are easily obtained.


\section{Fits and Predictions}\label{sec:fp}





In the previous section, the significance of the flavor-spin
interactions in establishing the correct level ordering for the Roper
and N$^*$(1535) resonances was pointed out. Here we will focus on the
effects of flavor-spin interactions in the case where SU$(3)_F$ is
broken both by the strange quark mass and by the flavor-spin
interactions when $C_{SI}$, $C_{47}$, and $C_8$ in Eq.~(\ref{eq:seven})
are unequal.   We consider three quark systems first to determine
the relevant parameters.

We obtain the values for coefficients in Eq.~(\ref{eq:seven}) by fitting the mass spectrum of the low-lying octet and decuplet baryons. For a specific $q^3$ state the mass $M$ is given by 
\begin{equation}						\label{eq:eight}
M = M_{0}^{(3)} + x_1 C_{SI} + x_2 C_{47} + x_3 C_8 + n_s \Delta m_s  
								\ ,
\end{equation}

\noindent where $M_0^{(3)}$ is a base mass, $x_1$, $x_2$, and $x_3$ are matrix elements of the operators in Eq.~(\ref{eq:seven}), $n_s$ is the number of strange quarks, and $\Delta m_s$ is the mass increase due to the presence of a strange quark.

\begin{table}[ht]

\begin{tabular}{lrrrr}
\hline\hline
State\qquad\qquad &  \qquad $x_1$ \qquad & \qquad $x_2$\qquad & \qquad $x_3$ \qquad & \qquad $n_s$ \qquad \\
\hline     
$N$  \qquad\qquad          &  \qquad $-15$ \qquad & \qquad $0$ \qquad   & \qquad $1$ \qquad  & \qquad 0 \qquad \\
$\Delta$\qquad\qquad   &  \qquad $-3$ \qquad  & \qquad 0 \qquad   & \qquad $-1$ \qquad & \qquad 0 \qquad \\
$\Lambda$\qquad\qquad  &  \qquad $-9$ \qquad  & \qquad $-6$ \qquad  & \qquad 1 \qquad  & \qquad 1 \qquad \\
$\Sigma$\qquad\qquad   &  \qquad $-1$ \qquad  & \qquad $-10$ \qquad & \qquad $-3$ \qquad & \qquad 1 \qquad \\
$\Sigma^*$\qquad\qquad &  \qquad $-1$ \qquad  & \qquad $-4$ \qquad  & \qquad 1 \qquad  & \qquad 1 \qquad \\
$\Xi$\qquad\qquad &  \qquad 0 \qquad   & \qquad $-10$ \qquad & \qquad $-4$ \qquad & \qquad 2 \qquad \\
$\Xi^*$\qquad\qquad &  \qquad 0 \qquad & \qquad $-4$ \qquad  & \qquad 0 \qquad  & \qquad 2 \qquad \\
$\Omega$\qquad\qquad   &  \qquad 0 \qquad   & \qquad 0 \qquad   & \qquad $-4$ \qquad & \qquad 3 \qquad \\
\hline\hline 
\end{tabular}

\caption{Numerical coefficients for Eq.~(\ref{eq:eight}).}

\label{massprediction}
\end{table}

We fit $M_0^{(3)}$, $\Delta m_s$, $C_{SI}$, $C_{47}$ and $C_8$ to the
well-known masses of the baryons listed in Table~\ref{masscomparison}. 
The experimental masses given are isospin averages.  The results are:
\begin{eqnarray}
M_0^{(3)} &=& 1340.5 \pm 5.3 {\rm\ MeV}, \quad
	\Delta m_s = 136.3 \pm 2.5 {\rm\ MeV} \nonumber \\
C_{SI} &=& 28.2 \pm 0.5 {\rm\ MeV,} \quad
	C_{47} = 20.7 \pm 0.5 {\rm\ MeV,} \quad
C_8 = 19.7 \pm 1.2 {\rm\ MeV}
 \ .
\end{eqnarray}

\noindent An error of 5 MeV is assumed  for each of the baryon masses, to take into account theoretical uncertainties.  Thus, moving any of the parameters to the edge of the quoted error limits changes the predicted baryon masses by about 5 MeV.
With these parameters, and the Roper fixed at 1440~MeV, the S$_{11}$ mass is predicted to be 1526~MeV.

\begin{table}[ht]
\begin{tabular}{lcc}
\hline\hline
State\qquad\qquad &  \qquad Experimental Mass (MeV) \qquad & \qquad Predicted Mass (MeV) \qquad \\
\hline 
$N$ \qquad\qquad        &  \qquad $939$ \qquad  & \qquad $937$ \qquad   \\
$\Delta$\qquad\qquad   &  \qquad $1232$ \qquad  & \qquad $1236$ \qquad    \\
$\Lambda$\qquad\qquad  &  \qquad $1116$ \qquad  & \qquad $1119$ \qquad  \\
$\Sigma$\qquad\qquad   &  \qquad $1193$ \qquad  & \qquad $1183$ \qquad  \\
$\Sigma^*$\qquad\qquad &  \qquad $1385$ \qquad  & \qquad $1386$ \qquad  \\
$\Xi$\qquad\qquad      &  \qquad $1318$ \qquad  & \qquad $1327$ \qquad  \\
$\Xi^*$\qquad\qquad    &  \qquad $1533$ \qquad  & \qquad $1530$ \qquad  \\
$\Omega$\qquad\qquad   &  \qquad $1672$ \qquad  & \qquad $1670$ \qquad  \\
\hline\hline 
\end{tabular}
\caption{Fit to the low-lying octet and decuplet baryon masses, using the predictions given by Eq.~(\ref{eq:seven}) and Table~\ref{massprediction}.}

\label{masscomparison}
\end{table}




Applying the same approach to the pentaquark antidecuplet, we obtain a mass $M$ for each state given by:
\begin{equation}						\label{eq:ten}
M = M_{0}^{(5)} + x_1 C_{SI} + x_2 C_{47} + x_3 C_8 
	+ n_s^{eff} \Delta m_s  
								\ .
\end{equation}

\noindent $M_0^{(5)}$ is the base mass for 5-quark bound states and
should be different from $M_0^{(3)}$ found earlier.  The values for
model parameters given in Eq.~(\ref{eq:eight}) can change in going from
q$^3$ system to $q^{4}\bar q$ system. We anticipate that the largest
change in the model parameters will occur in $M_0$, while we expect the
other parameters to have a less marked dependence on the number of quarks.
Therefore we proceed by eliminating $M_0^{(5)}$ from the mass formula
by the use of the experimentally measured mass of the $\Theta^+$,
$M_{\Theta}$=1542 MeV~\cite{nakano,barmin,stepanyan,barth,kubarovsky}.  The number $n_s^{eff}$, is the expectation value of the number of strange quarks plus strange antiquarks in each state, taking due account of hidden strangeness components, which were shown to be significant in~\cite{us}.  The necessary matrix elements may be evaluated using the pentaquark maximally symmetric flavor-spin wave function, which can be written as\footnote{
The four-quark part of this state is totally antisymmetric, as it should be.  A diquark-diquark state, such as in~\cite{jaffew}, has antisymmetry within each diquark, but antisymmetry when exchanging quarks between different diquarks is not enforced.  This can be viewed as an approximation that is valid if the diquarks are much smaller than the overall state.  In a absence of a mechanism that compresses the diquarks, a diquark-diquark state violates Fermi-Dirac statistics.
}
\begin{equation}
\left| (\overline{\mathbf{10}},1/2) \right\rangle
= {1\over \sqrt 2} 
                   \left\vert (\bar {\bf 3},0)(\bar {\bf 3},0)
                   \right\rangle_{\bar {\bf 6},0}+
                   {1\over \sqrt 2}
                   \left\vert ({\bf 6},1)({\bf 6},1)
                   		\right\rangle_{\bar {\bf 6},0}  \ ,
\label{psi-4q}
\end{equation}

\noindent where the pair of numbers in parentheses refer to the flavor and spin.  On the right hand side, the first (second) pair of numbers refers to the first (second) pair of quarks, and the quantum numbers of the antiquark $(\bar{\mathbf 3},1/2)$ are the same in each term and have been suppressed.  The numerical values of the matrix elements in Eq.~(\ref{eq:ten}) are given in Table~\ref{herry}.

\begin{table}[ht]

\begin{tabular}{lcccc}
\hline\hline
State\qquad\qquad &  \qquad $x_1$ \qquad & \qquad $x_2$\qquad & \qquad $x_3$ \qquad & \qquad $n_s^{eff}$ \qquad 
\\
\hline     
$\Theta$  \qquad\qquad          &  \qquad $-30$ \qquad & \qquad $0$ \qquad   & \qquad $2$ \qquad  & \qquad 1 \qquad 
\\
$N_5$ \qquad\qquad   &  \qquad $-20$ \qquad  & \qquad $-8$ \qquad   & \qquad $0$ \qquad & \qquad $4\over 3$ \qquad 
\\
$\Sigma_5$\qquad\qquad  &  \qquad $-{31\over 3}$ \qquad  & \qquad $-{44\over 3}$ \qquad  & \qquad $-3$ \qquad  & \qquad $5\over 3$ \qquad 
\\
$\Xi_5$\qquad\qquad   &  \qquad $-1$ \qquad  & \qquad $-20$ \qquad & \qquad $-7$ \qquad & \qquad 2 \qquad \\
\hline\hline 
\end{tabular}

\caption{Numerical coefficients for Eq.~(\ref{eq:ten}).}

\label{herry}
\end{table}

Using the wave function given by Eq.~(\ref{psi-4q}), and the mass
formula expressed in Eq.~(\ref{eq:ten}), we find the following masses
for the members of the baryon antidecuplet: $M(N_5)=1665$~MeV,
$M(\Sigma_5)=1786$~MeV and $M(\Xi_5)=1906$~MeV. To complete our
predictions, we use the predicted mass spectrum and SU$(3)_F$ symmetry
for the decay matrix elements to estimate widths of the decay modes of
the highest isospin members of the antidecuplet. Table~\ref{decaytable}
lists our predictions.

\begin{table}[ht]

\begin{tabular}{lcclcc}
\hline\hline
Decay\qquad\qquad & \qquad $|A/A_0|^2$ \qquad & \qquad $\Gamma/\Gamma_0$ \qquad\qquad & Decay \qquad\qquad &
\qquad $|A/A_0|^2$ \qquad & $\Gamma/\Gamma_0$ \\
\hline 
$\Theta^+\rightarrow p K^0$ & $1$ & $0.97$ & $\Sigma_5^+ \rightarrow \Sigma^+ \eta$ & $1/2$  &  $0.13$ \\
$p_5\rightarrow \Lambda K^+$ & $1/2$ & $0.15$ & $\Sigma_5^+ \rightarrow \Lambda \pi^+$ & $1/2$ &  $2.63$ \\
$p_5\rightarrow p \eta$ &  $1/2$ & $1.10$ & $\Sigma_5^+ \rightarrow  p \bar K^0$ & $1/3$ &  $1.86$ \\
$p_5\rightarrow \Sigma^+ K^0$ & $1/3$ & $-$ & $\Sigma_5^+ \rightarrow \Sigma^+ \pi^0$ & $1/6$ &  $0.63$\\
$p_5\rightarrow \Sigma^0 K^+$ & $1/6$ &   $-$ & $\Sigma_5^+ \rightarrow \Sigma^0 \pi^+$ & $1/6$ &  $0.61$ \\
$p_5\rightarrow n \pi^+$ & $1/3$ &  $2.48$ & $\Xi_5^+ \rightarrow \Xi^0 \pi^+$ & $1$ &  $3.23$ \\
$p_5\rightarrow p \pi^0$ & $1/6$ &  $1.25$ & $\Xi_5^+ \rightarrow \Sigma^+ \bar K^0$ & $1$ & $2.22$ \\
$\Sigma_5^+ \rightarrow \Xi^0 K^+$ & $1/3$ &   $-$ & &\\
\hline\hline 
\end{tabular}
\caption{SU(3) decay predictions for the highest isospin members of the 
positive parity antidecuplet. $A_0$ and $\Gamma_0$ are the amplitude and partial
decay width for $\Theta^+ \rightarrow n K^+$, respectively. Pentaquark masses are $1542$,
$1665$, $1786$, and $1906$~MeV, for the $\Theta^+$, $p_5$, $\Sigma_5$ and $\Xi_5$, respectively.}

\label{decaytable}

\end{table}

It should be stressed that we view the mass and decay predictions of the $\Xi_5$ states to be most reliable due to the absence of substantial mass mixing with nearby states.  While we provide predictions for the $N_5$ and $\Sigma_5$ for the sake of completeness, these may be subject to large corrections due to mixing with octet pentaquarks.  Whether such effects could be reliably evaluated is an interesting question which is beyond the scope of the present work.

\section{Conclusions}\label{sec:conc}

We have considered the possibility that the lightest strangeness one
pentaquark state is positive parity, with one unit of orbital angular
momentum.  In this case, it is possible to construct states with
totally symmetric spin-flavor wave functions. Spin-flavor exchange
interactions, if dominant, render these states lighter than any
pentaquark with all its constituents in the ground states.  We assume
such spin-flavor exchange interactions in an effective theory,
including flavor SU(3) breaking effects in operator coefficients and in
the quark masses.  The general form of these multi-quark interactions
is consistent with a number of possible models of the underlying
dynamics, including pseudoscalar meson exchange, skyrmions, and
instanton-induced effects.  In our approach, however, we need not
commit ourselves to any specific dynamical picture.  We believe that
the theoretical uncertainty in using such a streamlined (yet pragmatic)
approach is no greater than the spread in predictions between different
specific models.  Use of effective spin-flavor exchange interactions is
well motivated given its success in explaining the lightness of the
Roper resonance relative to the negative parity N(1535), as we
demonstrated in Section~\ref{sec:fw}.  Simple quark models without dominant
spin-flavor exchange interactions simply get the ordering of these
states wrong. Fitting our operator coefficients, a mean multiplet mass,
and a strangeness mass contribution to the masses of the ground state
octet and decuplet baryons, we then predict mass splittings in the
parity even pentaquark antidecuplet.  In particular,
our approach allows us to predict the mass of the strangeness $-2$
cascade states at $1906$~MeV, with a full width approximately $2.8$
times larger than that of the $\Theta^+$.   The cascade states do not
mix with any other pentaquarks of comparable mass, which makes these
prediction particularly robust.  Discovery of cascade pentaquarks
around $1906$~MeV would therefore provide an independent test of the
importance of spin-flavor exchange interactions in the breaking of the
approximate SU(6) symmetry of the low-lying hadron spectrum.


\begin{acknowledgments}
We thank Leonid Glozman and Josef Speth for useful comments.  We thank the NSF for
support under Grant Nos.\ PHY-0140012, PHY-0243768 and PHY-0245056.
\end{acknowledgments}


\end{document}